\documentclass[11pt,twoside]{article}

\usepackage{asp2006}
\usepackage{epsf}
\usepackage{lscape}
\usepackage{natbib}

\begin{document}

\title{Modeling the X-ray fractional variability spectrum of Active
  Galactic Nuclei using multiple flares}

\author{R.~W.~Goosmann, M.~Dov\v{c}iak, V.~Karas}
\affil{Astronomical Institute, Academy of Sciences, Prague, Czech Republic}   
\author{B.~Czerny}
\affil{Copernicus Astronomical Center, Warsaw, Poland}   
\author{M.~Mouchet}
\affil{Laboratoire ApC, Universit\'e Denis Diderot, Paris, France}
\author{G.~Ponti}
\affil{Dipartimento di Astronomia, Universit\`a di Bologna, Bologna, Italy}   

\begin{abstract}
Using Monte-Carlo simulations of X-ray flare distributions across the
accretion disk of active galactic nuclei (AGN), we obtain modeling
results for the energy-dependent fractional variability
amplitude. Referring to previous results of this model, we illustrate
the relation between the shape of the point-to-point fractional
variability spectrum, $F_{\rm pp}$, and the time-integrated spectral
energy distribution, $F_{\rm E}$. The results confirm that the
spectral shape and variability of the iron K$\alpha$ line are
dominated by the flares closest to the disk center.
\end{abstract}

The fractional variability spectrum of AGN describes the
variability properties as a function of photon energy \citep[see
e.g.][]{edelson2002, vaughan2003}. In \citet{goosmann2006} we present modeling
of AGN fractional variability spectra for distributions of magnetic
flares co-rotating with the accretion disk. Using a Monte-Carlo
method, we sample the time evolution of the flare distribution for
different choices of the global parameters. These determine the
radial distribution of the flare number and luminosity across the
disk, the individual flare life times, the average X-ray luminosity of
the object, the mass of the black hole, and its spin. By using a
ray-tracing method, the computations include general relativistic and
Doppler corrections.

For this proceedings note we present an extension of our work by
emphasizing the connection between $F_{\rm pp}$ and $F_{\rm E}$ for a
given model setup. We consider parameter sets that only differ in the
radial distribution of the individual flare luminosity, ruled by the
parameter $\beta$. All other parameters are set as for our best fit to
the $F_{\rm pp}$ variability spectrum of MCG-6-30-15, which was
computed from a 95 ksec {\it XMM-Newton} observation \citep[for
details see][]{goosmann2006}.

\begin{figure}[t]
 \centering
 \epsfxsize=12.6cm
 \epsfbox{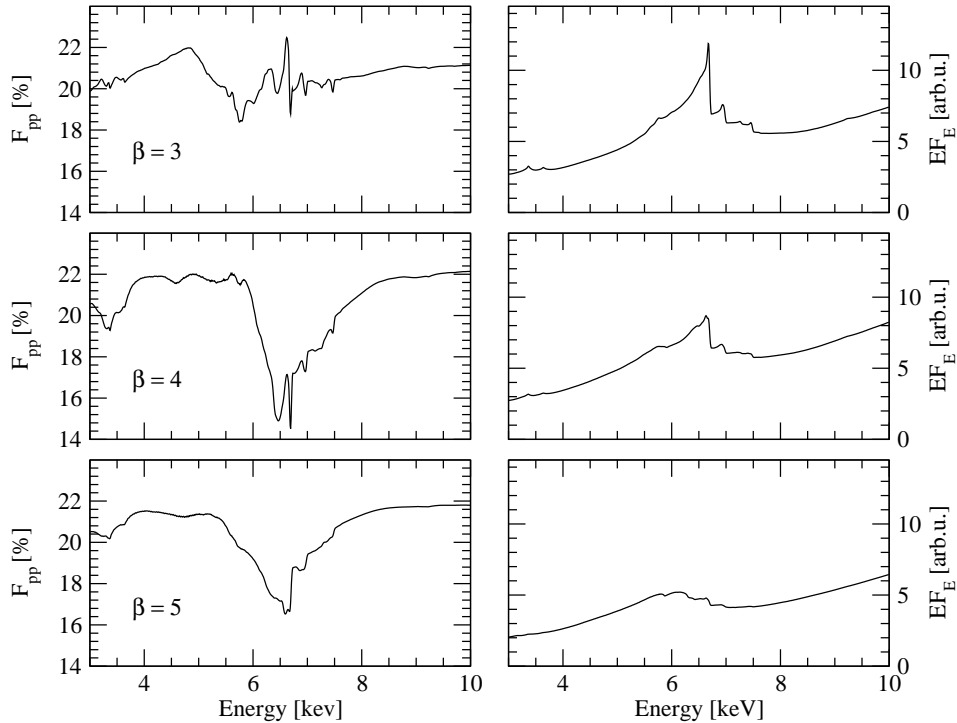}
 \caption{Point-to-point fractional variability $F_{\rm pp}$ (left)
 and the corresponding time-integrated spectra (right) for flare
 distributions with different $\beta$. The $F_{\rm pp}$ spectra are
 all normalized to the same variability level.\label{fig:res}}
\end{figure}

In Fig.~\ref{fig:res} we show results for $F_{\rm pp}$ and $F_{\rm E}$
with $\beta =$ 3, 4 and 5. The normalization (but not the shape) of
$F_{\rm pp}$ can vary for the same parameter set within the limits of
Monte-Carlo statistics. Therefore, we normalize the $F_{\rm pp}$
spectra in Fig.~\ref{fig:res} to the variability level of $\sim 20\%$
as obtained for our best fit to MCG-6-30-15. Thus, we only investigate
the shape of $F_{\rm pp}$ and not the normalization.

For $\beta = 3$, when the radial profile of the X-ray luminosity is roughly
proportional to the radial profile of the disk luminosity, the variations of
$F_{\rm pp}$ with energy are relatively low. For higher values of $\beta$ the
variability is significantly depressed around the iron K$\alpha$ line, and the
iron line profile is more strongly smeared out. Hence, if the X-ray luminosity
is due to individual, localized flares, and if their luminosity rises more
strongly than the disk luminosity toward the center, then we expect not only a
broadened iron-line profile, but also a significant depression of the spectral
variability across the line. Note that such a strong rise of the disk
irradiation toward the center is predicted by the light-bending models
\citep[see e.g.][]{suebsuwong2006}.

\acknowledgements We are grateful to A.-M. Dumont and
A. R{\'o}{\.z}a{\'n}ska for their help with computing the local flare
spectra used in this model.

\end{document}